\documentclass[preprint,showpacs,preprintnumbers,amsmath,amssymb]{revtex4}

\usepackage{color,vmargin,epsfig,rotating,graphicx,subfigure}
\usepackage{amsmath,amssymb,amscd}
\usepackage{textcomp}
\usepackage[latin1]{inputenc}

\setpapersize{A4}

\begin{document}

\title{Near-field thermal imaging of nano-structured surfaces}

\author{A. Kittel, U. F. Wischnath, J. Welker, O. Huth, F. R\"uting, and
	S.-A. Biehs}

\affiliation{Institut f\"ur Physik, Carl von Ossietzky Universit\"at,
        D-26111 Oldenburg, Germany}

\date{October 10, 2008}

\pacs{44.40.+a, 78.68.+m, 05.40.-a, 41.20.Jb}

\begin{abstract}
We show that a near-field scanning thermal microscope, which essentially
detects the local density of states of the thermally excited electromagnetic
modes at nanometer distances from some material, can be employed for nanoscale
imaging of structures on that material's surface. This finding is explained
theoretically by an approach which treats the surface structure perturbatively.
\end{abstract}

\maketitle

%%%%%%%%%%%%%%%%%%%%%%%%%%%%%%%%%%%%%%%%%%%%%%%%%%%%%%%%%%%%%%%%%%%%%%%%%%%%%%%%

%%%%%%%%%%%%%%%%%%%%%%%%%%%%%%%%%%%%%%%%%%%%%%%%%%%%%%%%%%%%%%%%%%%%%%%%%%%%%%%%
%
% introduction
%
%%%%%%%%%%%%%%%%%%%%%%%%%%%%%%%%%%%%%%%%%%%%%%%%%%%%%%%%%%%%%%%%%%%%%%%%%%%%%%%%

It is well known that a dielectric body kept at constant temperature is
surrounded by the fluctuating thermal electromagnetic field generated by
fluctuations inside the material. These fluctuating fields can be divided
into propagating and evanescent modes, {\em i.e.\/}, radiating and
nonradiating modes. The familiar radiating modes can be described by means
of the Kirchhoff-Planck law of thermal radiation. Furthermore, the
evanescent modes can also contribute to the radiative heat transfer due to
photon tunneling, if a second dielectric body with a different temperature
is brought close to the first one. This additional contribution can exceed
the radiative heat flux between two black bodies by several orders of
magnitudes if the distance between them is much smaller than the dominant
thermal wavelength.

A theoretical description within the framework of Rytov's fluctuational
electrodynamics~\cite{RytovEtAl89} of the radiative heat transfer between two
closely spaced semi-infinite bodies with parallel surfaces was given by Polder
and van Hove~\cite{PolderVanHove71}. Early experimental observations of
the additional contribution of the evanescent modes in a geometry of two
parallel plates were made by Hargreaves~\cite{C.M.Hargreaves_1969} and by
Domoto {\itshape et al.}~\cite{DomotoEtAl70}. In a quite recent experiment
by Hu {\itshape et al.} with precision glass optical flats~\cite{HuEtAl08},
good agreement with the prediction of Polder and van Hove has been found for
a distance of 1.6~\textmu m. It is rather difficult to reach shorter distances
in such experiments with plates, because these plates have to be kept
perfectly parallel. Moreover, the achievable minimum distances are limited
by the surface roughness. These restrictions are overcome by a near-field
scanning thermal microscope (NSThM)~\cite{MuellerHirsch1999,A.Kittel_2005}.
The STM tip of such an instrument also serves as a thermocouple, so that
the radiative heat transfer between a cooled sample and this probe can be
measured directly for distances ranging from 1~nm up to the resolution limit
of the apparatus, which currently lies in the range of some
10~nm~\cite{WishnathEtAl08}. Hence, the NSThM acts in a regime of
probe-sample distances in which simple local macroscopic models for the
radiative near-field heat transfer~\cite{Dorofeyev98,JPendry99,MuletEtAl01,
A.I.VolokitinB.N.J.Persson01,DedkovKyasov07} based on Rytov's theory may
become questionable~\cite{C.Henkel_K.Joulain_2006,I.A.Dorofeyev_2007}.
In this letter, we demonstrate that an NSThM provides images of
a nanoscale structured surface. Despite the above objection, the measured signals can
at least qualitatively be understood within a simple local approach,
describing the sensor tip by means of a dipole model and the sample's surface
by perturbation theory~\cite{BiehsEtAl2008}.

%%%%%%%%%%%%%%%%%%%%%%%%%%%%%%%%%%%%%%%%%%%%%%%%%%%%%%%%%%%%%%%%%%%%%%%%%%%%%%%%%%%%%%%%%%%%%%%%%%%%%%%%%%%%%%%%
%
% experimental setup and scanning mode
%
%%%%%%%%%%%%%%%%%%%%%%%%%%%%%%%%%%%%%%%%%%%%%%%%%%%%%%%%%%%%%%%%%%%%%%%%%%%%%%%%%%%%%%%%%%%%%%%%%%%%%%%%%%%%%%%%

The experiments were conducted with an NSThM developed~\cite{MuellerHirsch1999,
A.Kittel_2005} and refined~\cite{WishnathEtAl08} in our group. It is based on
a commercial variable temperature ultra-high vacuum scanning tunneling
microscope (VT-UHV STM). In order to measure the near-field contribution
to the heat flux, the STM tip has been functionalized to act as a microscopic
thermometer. This has been achieved by implementing a gold-platinum
thermocouple in coaxial configuration at the very end of the tip. The
fabrication of these thermocouple tips is described in detail
in Ref.~\cite{WishnathEtAl08}. At the foremost end, the tip consists of a
platinum wire with a tip radius of less than 50~nm, protruding about
500~nm from a glass capillary and covered by a thin gold film (few nanometers thick), as sketched in
Fig.~1(a). A micrograph taken with a scanning electron microscope
(SEM) is shown in Fig.~1(b).

During a measurement the sample is cooled down to about 110~K, whereas
the tip holder remains at ambient temperature (about 293~K). Due to near-field heat
transfer, the very end of the tip becomes slightly colder than its backside.
This small temperature difference is the source of the detected
thermovoltage~$V_{\rm th}$. Knowing both the tip's thermal resistance
and the Seebeck coefficient of the thermocouple, even an absolute
measurement of the near-field heat current between probe and sample is
possible~\cite{WishnathEtAl08}. Here we merely consider the thermovoltage
itself, which is proportional to the heat flux.

Figure~2(a) displays an STM micrograph of a gold surface
fabricated on mica, which exhibits some structure. During the scan performed at
constant distance (constant-current mode: $I_t=2\,{\rm nA}$, $V_t=0.5\,{\rm V}$), the
thermovoltage $V_{\rm th}$ has been detected at each point. The resulting
distribution of the thermovoltage is shown in Fig.~2(b).
Obviously, the heat flux across the vacuum gap between probe and sample
reflects the topography, even though the probe-sample distance is kept
constant. However, it does not give quite the same information as the 
topographic image; in particular, it tends to emphasize the structure's edges. This is
in itself an important finding: The NSThM provides nanoscale thermal surface
images.

%%%%%%%%%%%%%%%%%%%%%%%%%%%%%%%%%%%%%%%%%%%%%%%%%%%%%%%%%%%%%%%%%%%%%%%%%%%%%%%%%%%%%%%%%%%%%%%%%%%%%%%%%%%%%%%%
%
% theoretical model
%
%%%%%%%%%%%%%%%%%%%%%%%%%%%%%%%%%%%%%%%%%%%%%%%%%%%%%%%%%%%%%%%%%%%%%%%%%%%%%%%%%%%%%%%%%%%%%%%%%%%%%%%%%%%%%%%%

In what follows we explain this finding by comparing the measured signals
with results obtained from a theoretical model describing the probe's tip as
a metallic sphere, and considering its electric and magnetic dipole moments.
Such a model leads to a general expression for the radiative heat transfer
between the sensor tip and an arbitrary body~\cite{Dorofeyev08,ChapuisEtAl08}.
The geometry and materials properties of that body are incorporated via its
electric and magnetic local density of states (LDOS)~\cite{JoulainEtAl03}
evaluated at the point of measurement, {\em i.e.\/}, at the point inside the
tip where the dipole moments are located. The LDOS above a surface with a given
structure has recently been calculated by means of a perturbative approach up
to first order in the surface profile~\cite{BiehsEtAl2008}. It has been shown
that the signal detected by a spherical probe when scanning a surface with
some topographical structure coincides reasonably well with the sample's LDOS,
evaluated at the dominant thermal frequency. For metals such as gold, the
magnetic part of the LDOS dominates in the distance regime of
interest~\cite{ChapuisEtAl08}. Therefore, in order to qualitatively compare
the experimental data with the theory, it suffices to calculate the magnetic
LDOS at the dominant thermal frequency. While a full-fledged theoretical
treatment would require significantly more effort, this coarse approach
can be expected to capture the significant features.

As detailed in Ref.~\cite{BiehsEtAl2008}, this description is subject to
several restrictions. Firstly, the dipole approximation is strictly valid
only for probe-sample distances much greater than the tip radius (about 50$\,$nm),
whereas for smaller distances higher moments have to be considered.
Secondly, the perturbative approach provides a viable approximation for
observation distances much greater than the maximum height variation of the
surface profile (about 5$\,$nm). Thirdly, at distances much smaller than the mean free path (about 20$\,$nm)
of the conduction electrons, nonlocal effects have to be taken into account.
All three issues are relevant for our experiment. Nonetheless, we show that
the signals obtained with the NSThM can at least qualitatively be understood
on the basis of our simple approach, thus clarifying the physical principles
underlying near-field thermal imaging.

%%%%%%%%%%%%%%%%%%%%%%%%%%%%%%%%%%%%%%%%%%%%%%%%%%%%%%%%%%%%%%%%%%%%%%%%%%%%%%%%%%%%%%%%%%%%%%%%%%%%%%%%%%%%%%%%
%
% discussion of the results
%
%%%%%%%%%%%%%%%%%%%%%%%%%%%%%%%%%%%%%%%%%%%%%%%%%%%%%%%%%%%%%%%%%%%%%%%%%%%%%%%%%%%%%%%%%%%%%%%%%%%%%%%%%%%%%%%%

We calculated the zeroth- and first-order magnetic LDOS in a constant distance
of 9~nm above the surface of the sample via a discrete Fourier transform of the sample 
profile directly taken from the STM data shown in Fig.~2 without any smoothing. 
According to Ref.~\cite{BiehsEtAl2008}, the LDOS has to be 
evaluated at the dominant thermal frequency of the tip $\omega_{\rm th} = 10^{14}\,{\rm rad}\,{\rm s}^{-1}$, as 
corresponding to 293~K. The result of the calculation is shown in Fig.~3(a). 
In order to account for the proportionality between thermovoltage and heat
flux on the one hand, and for the approximate proportionality between LDOS
and heat flux on the other, we fitted the thermovoltage data to the LDOS performing a least-square fit. 
Scaling and shifting the thermovoltage data in Fig~3(b) accordingly, the overall 
agreement of their main features with the LDOS becomes obvious. We have to emphasize that 
it is not clear {\itshape a priori} which distance should be used for the calculation of 
the LDOS, since the radius of the foremost part of the probe's interaction
region is unknown. Our choice of 9~nm appears plausible, and yields good
agreement between the LDOS and the NSThM signal.  

There is no strict one-to-one coincidence between the NSThM data and the
LDOS. In view of the many simplifications underlying the theoretical description,
some deviations have to be expected. However, the model actually does
reproduce the main experimental trend: The NSThM-signal obtained in the
constant-current mode roughly corresponds to the inverted profile function.
In Fig.~4 we show two linescans taken from the 
scaled thermovoltage data, and from the theoretically obtained LDOS in Fig.~3,
for the horizontal and vertical line indicated by the arrows in Fig.~2.  
In both cases a good qualitative agreement between the experimental
and the theoretical data can be observed. However, for the horizontal linescan the LDOS
and the thermovoltage data appear to be shifted against each other in the $x$-direction, 
suggesting that the centers of interaction for the STM and the NSThM are shifted in 
that direction. This finding might be explained by the shape of the foremost part of 
the tip, which is not perfectly axially symmetric.    

%%%%%%%%%%%%%%%%%%%%%%%%%%%%%%%%%%%%%%%%%%%%%%%%%%%%%%%%%%%%%%%%%%%%%%%%%%%%%%%%%%%%%%%%%%%%%%%%%%%%%%%%%%%%%%%%
%
% summary
%
%%%%%%%%%%%%%%%%%%%%%%%%%%%%%%%%%%%%%%%%%%%%%%%%%%%%%%%%%%%%%%%%%%%%%%%%%%%%%%%%%%%%%%%%%%%%%%%%%%%%%%%%%%%%%%%%

To conclude, we have demonstrated that an NSThM operated at a
constant distance provides a thermal near-field image exhibiting nanoscale features of isothermal, structured
surfaces. The measured signal qualitatively follows the local density of
states of the thermal electromagnetic field above the surface, obtained by first-order perturbation theory
in the profile function and evaluated at the dominant thermal frequency,
neglecting nonlocal effects. More detailed, quantitative comparisons with
the predictions of a refined model are desirable for further exploring
the possibilities opened up by near-field thermal imaging.

\acknowledgments
This work was supported by the DFG through grant No.\ KI 438/8-1.

\newpage

%%%%%%%%%%%%%%%%%%%%%%%%%%%%%%%%%%%%%%%%%%%%%%%%%%%%%%%%%%%%%%%%%%%%%%%%%%%%%%%%%%%%%%%%%%%%%%%%%%%%%
%
% Bibliography
%
%%%%%%%%%%%%%%%%%%%%%%%%%%%%%%%%%%%%%%%%%%%%%%%%%%%%%%%%%%%%%%%%%%%%%%%%%%%%%%%%%%%%%%%%%%%%%%%%%%%%%

\newpage

\subsection*{Figures}

\begin{figure}[Hhbt]
  \centering
    \includegraphics[width=12cm]{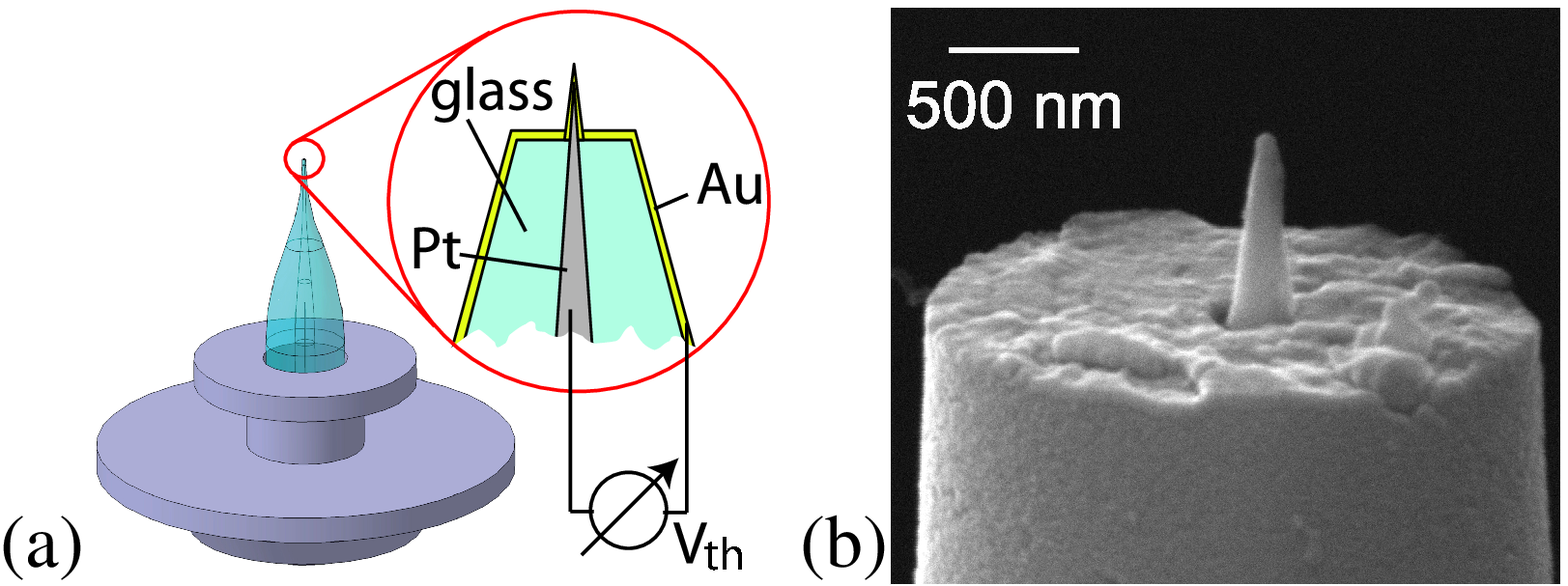}
    \caption{
    (a) Schematic drawing of the tip in its holder, with blown-up
    cross section of the tip's very end, displaying the glass capillary,
    the platinum wire, and the gold coating which form the thermocouple.
    (b) SEM image of a typical NSThM tip. The platinum wire protrudes from
    the center of the glass mantle. Both are covered by a gold film.
    }
\end{figure}

\vspace{2cm}

\begin{figure}[Hhbt]
  \centering
    \includegraphics[width=12cm]{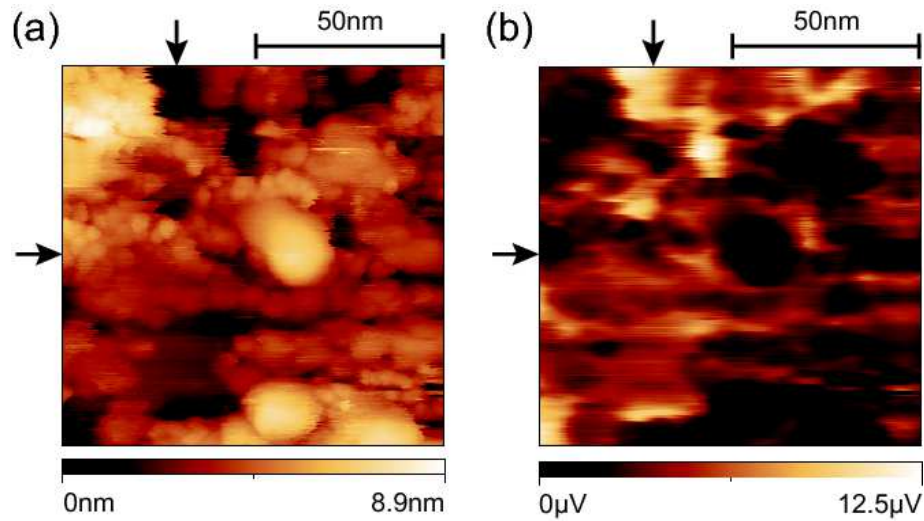}
    \caption{
    (a) STM-topography of a gold surface on mica. (b) Spatial
    distribution of the local thermovoltage measured by the thermocouple.
    The temperature of the sample is 110~K, that of the probe 293~K.
    }
\end{figure}

\vspace{2cm}

\begin{figure}[Hhbt]
  \centering
    \includegraphics[width=12cm]{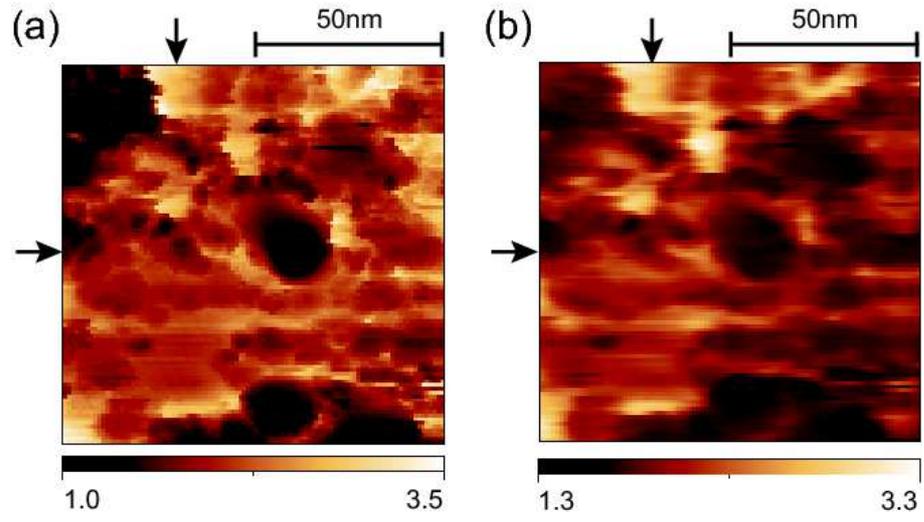}
    \caption{
     (a) Numerically calculated LDOS in $10^6\,{\rm m}^{-3}\,{\rm s}$ at a constant 
     distance of $9\,{\rm nm}$ above the two-dimensional topography directly extracted from 
     the STM data in Fig.~2. (b) A plot of the thermovoltage data (in arbitrary units) from Fig.~2 rescaled as described in the text.
    }
\end{figure}

\vspace{2cm}

\begin{figure}[Hhbt]
  \centering
    \includegraphics[width=12cm]{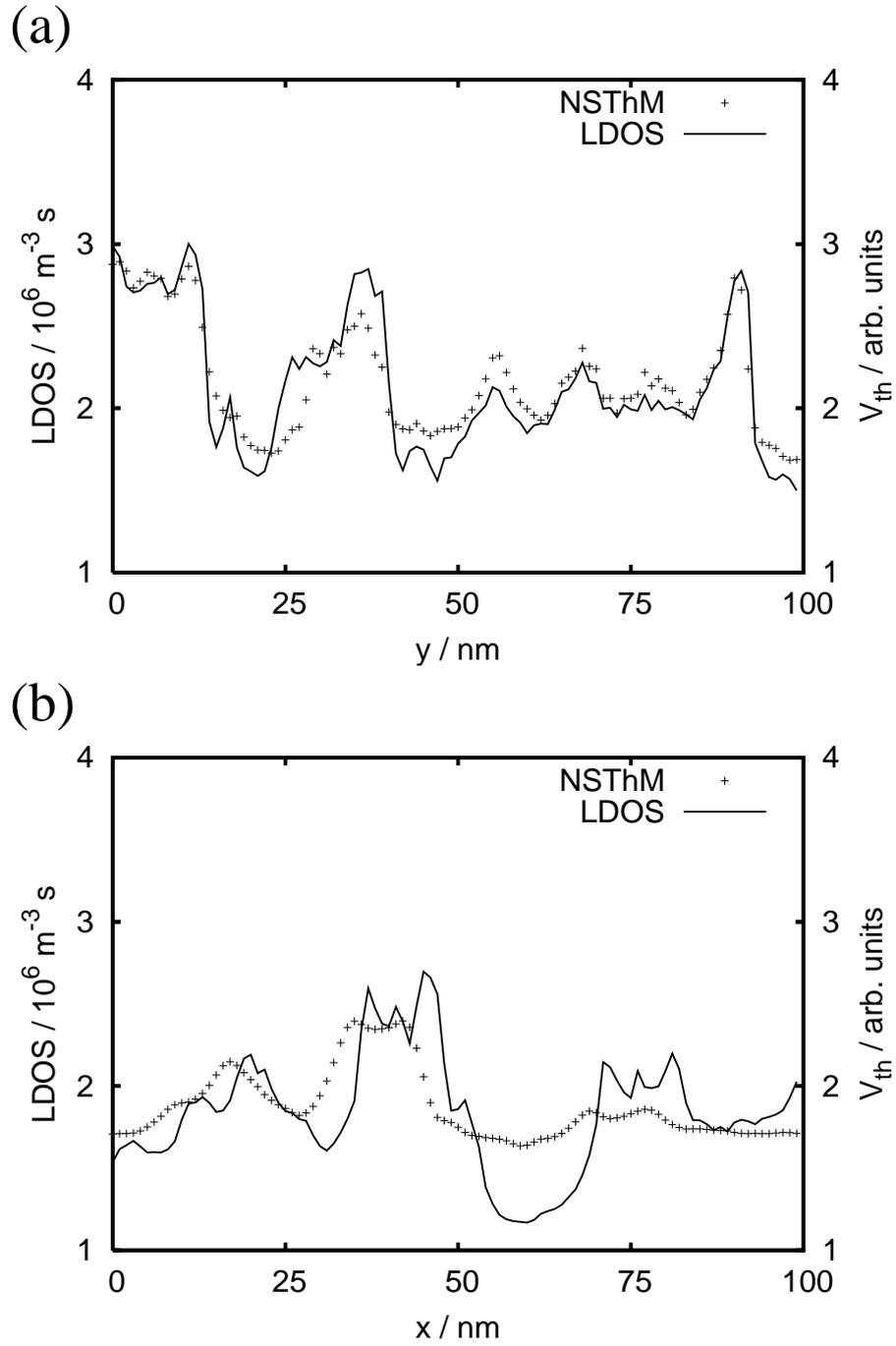}
    \caption{
    (a) LDOS at a constant distance of 9~nm above the surface profile in comparision to the
    thermovoltage data, for the linescan indicated by the vertical arrow
    in Fig.~3. (b) LDOS at a constant distance of 9~nm above the profile in comparision to the
    thermovoltage data, for the line scan indicated by the horizontal arrow in Fig.~3.
\
    }
\end{figure}

\end{document}